\documentclass[12pt]{article}
\usepackage{amssymb,amsmath,epsfig}
\begin{document}
\title{\bf Anisotropic Universe in $f(\mathcal{G},\textit{T})$ Gravity}

\author{M. Farasat Shamir\thanks{farasat.shamir@nu.edu.pk}\\\\
Department of Sciences and Humanities, \\National University of
Computer and Emerging Sciences,\\ Lahore Campus, Pakistan.}

\date{}

\maketitle
\begin{abstract}
This paper is devoted to investigate the recently introduced $f(\mathcal{G},\textit{T})$ theory of gravity, where $\mathcal{G}$ is the Gauss-Bonnet term, and ${\textit{T}}$ is the trace of the energy-momentum tensor. For this purpose, anisotropic background is chosen and a power law $f(\mathcal{G},\textit{T})$ gravity model is used to find the exact solutions of field equations.
In particular, a general solution is obtained which is further used to reconstruct some important solutions in cosmological contexts. The physical quantities like energy density, pressure, and equation of state parameter are calculated. A Starobinsky Like $f_2(\textit{T})$ model is proposed which is used to analyze the behavior of universe for different values of equation of state parameter. It is concluded that presence of term $\textit{T}$ in the bivariate function $f(\mathcal{G},\textit{T})$ may give many cosmologically important solutions of the field equations.
\end{abstract}


\section{Introduction}

Some important modifications of general relativity (GR) have been proposed in the last two decades.
The mostly discussed theories are $f(R)$ and $f(R,\textit{T})$ theories of gravity ($R$ is the Ricci scalar and $\textit{T}$ is the trace of energy-momentum tensor) have been treated most seriously \cite{5}-\cite{501}. However, recently $f(R,\textit{T})$ cosmology has been severely challenged \cite{ft}. On the other hand, modified Gauss-Bonnet (GB) gravity is another theory which has gained popularity in the last few years \cite{fG}. It is also known as $f(\mathcal{G})$ theory of gravity, where $f(\mathcal{G})$ is a generic function of GB invariant $\mathcal{G}$. GB term plays an important role as it may avoid ghost contributions and helps in regularizing the gravitational action \cite{Chiba}. Thus to save $f(\textit{T})$ theories, one may include $f(\mathcal{G})$ terms. A theory with a similar idea has been recently proposed named as $f(\mathcal{G},\textit{T})$ gravity \cite{sharif.ayesha}. Some interesting work has been done in the recent past using modified GB theories.

Anisotropic compact stars in modified $f(\mathcal{G})$ gravity have been discussed by Abbas et al. \cite{abbas}. Houndjo et al. \cite{Houndjo1} found the exact solutions of $f(\mathcal{G})$ field equations using cylindrical symmetry and it was concluded that
there existed seven families of exact solutions for three different forms of $f(\mathcal{G})$ models.
The exact cylindrically symmetric solutions of modified field equations recovered cosmic string space-time \cite{Houndjo2}. A more generalized version of GB gravity known as $f(R,\mathcal{G})$ gravity has also been discussed widely. Wu and Ma \cite{fRG1} investigated the spherically symmetric solutions at low energy where the weak-field and slow-motion limit of $f(R,\mathcal{G})$ gravity was developed. Laurentis et al. \cite{fRG2} argued cosmological inflation in $f(R,\mathcal{G})$ theory. Sharif and Ikram \cite{warm} examined warm inflation in $f(\mathcal{G})$ theory of gravity using scalar fields for the Friedmann-Robertson-Walker (FRW) universe model. Conserved quantities have been recently explored in FRW background using Noether symmetry \cite{noetherfG1}. Sharif and Fatima \cite{noetherfG2} discussed the role of GB term for the early and late time accelerating phases of the universe by considering two viable $f(\mathcal{G})$ models. Garcia et al. \cite{Garcia} explored energy conditions to prove the viability of some $f(\mathcal{G})$ gravity models. $f(R,\mathcal{G})$ gravity energy conditions have been recently explored where the WEC was used along with the recent estimated values of cosmological parameters to determine the viability of some specific choices of $f(R,\mathcal{G})$ gravity models \cite{Atazadeh}.

In the context of $f(\mathcal{G},\textit{T})$ gravity, Sharif and Ikram proved that the massive test particles followed non-geodesic lines of geometry due to the presence of extra force and examined the energy conditions for flat FRW universe \cite{sharif.ayesha}.
The same authors \cite{sharif.ayesha1} used reconstruction techniques to reproduce the cosmic evolution corresponding to de Sitter universe, power-law solutions and phantom/non-phantom eras in this theory. In a recent paper \cite{shamir.ahmad}, $f(\mathcal{G},\textit{T})$ theory of gravity was discussed using Noether symmetry approach. Two specific models were studied to determine the conserved quantities and it was concluded that the
well known deSitter solution could be reconstructed for some specific choice of $f(\mathcal{G},\textit{T})$ gravity model. In a recent paper, Sharif and Ikram \cite{sharif.ayesha2} studied the wormhole solutions using power law $f(\mathcal{G},\textit{T})$ gravity models and it was shown that traversable wormhole solutions were physically acceptable in $f(\mathcal{G},\textit{T})$ theory of gravity. In another work \cite{shamir.ahmad1}, Noether symmetry methodology has been used to study some cosmologically important $f(\mathcal{G},\textit{T})$ gravity models with anisotropic background. It is concluded that the specific models of modified GB gravity may be used to reconstruct $\Lambda$CDM cosmology without involving any cosmological constant. For some particular choices of $f(\mathcal{G},\textit{T})$ gravity models, it is anticipated that this theory may explain the late-time cosmic acceleration. Thus it seems interesting to explore further the modified $f(\mathcal{G},\textit{T})$ gravity. Moreover, in comparison with $f(R,\mathcal{G})$ gravity, the presence of the matter term $\textit{T}$ in the bivariate function $f(\mathcal{G},\textit{T})$ may give many solutions of the field equations and the theory may support the accelerated expansion of universe under certain conditions for the model under consideration.

In this paper, we are focussed to investigate the dynamics of $f(R)$ gravity with anisotropic background.
It is well known that the isotropic models are among the best choices to study
large scale structure of the universe. Moreover, according to the cosmological observations including
the cosmic microwave background (CMB) radiation, the current universe is isotropic.
However, it is believed that the early universe may not have been exactly uniform.
Also the local anisotropies that we observe today in galaxies and super clusters also motivate us to
model the universe with anisotropic background. Bianchi type models are among the simplest models
with anisotropic background. In particular, the investigation of Bianchi type universe in context of modified
theories is interesting. In this work, we are interested to explore $f(\mathcal{G}, \mathrm{\textit{T}})$ gravity using locally rotationally
symmetric (LRS) Bianchi type $I$ spacetime. We find the exact solutions of the LRS Bianchi type $I$ field equations in $f(\mathcal{G})$ theory of gravity.
In particular, a general solution with power law $f(\mathcal{G}, \mathrm{\textit{T}})$ gravity model is reported. The plan of paper is as follows:
Some basics of $f(\mathcal{G})$ gravity and field equations are discussed in section \textbf{2}. Section \textbf{3} is devoted to explore the exact solutions of modified field equations. Section \textbf{4} is used to reconstruct some important cosmological solutions. Final remarks are given in last section.

\section{Some Basics of $f(\mathcal{G},\mathrm{\textit{T}})$ Gravity}

The modified GB gravity is given by the action \cite{sharif.ayesha},
\begin{equation}\label{action}
\mathcal{A}= \frac{1}{2{\kappa}^{2}}\int d^{4}x
\sqrt{-g}[R+f(\mathcal{G},\mathrm{\textit{T}})]+\int
d^{4}x\sqrt{-g}\mathcal{L}_{M}.
\end{equation}
Here $\mathcal{G}$ and $\mathrm{T}$ denote the GB term and the trace of the energy-momentum tensor respectively, whereas $\mathcal{L}_{M}$ is the standard matter Lagrangian, $R$ is the Ricci Scalar, $g$ is the determinant of metric tensor and $\kappa$ is a coupling constant.
The field equations can be obtained by varying the action Eq.(\ref{action}) with respect to the metric tensor \cite{sharif.ayesha}
\begin{align}\label{4_eqn}
\begin{split}
&R_{\zeta\eta}-\frac{1}{2}g_{\zeta\eta}R=-[2Rg_{\zeta\eta}\nabla^{2}-2R\nabla_{\zeta}\nabla_{\eta}-4g_{\zeta\eta}R^{\mu\nu}\nabla_{\mu}\nabla_{\nu}-
4R_{\zeta\eta}\nabla^{2}+4R^{\mu}_{\zeta}\nabla_{\eta}\nabla_{\mu}+\\&
4R^{\mu}_{\eta}\nabla_{\zeta}\nabla_{\mu}+4R_{\zeta\mu\eta\nu}\nabla^{\mu}\nabla^{\nu}]f_{\mathcal{G}}(\mathcal{G},\mathrm{\textit{T}})+
\frac{1}{2}g_{\zeta\eta}f(\mathcal{G},\mathrm{\textit{T}})-
[\mathrm{\textit{T}}_{\zeta\eta}+\Theta_{\zeta\eta}]f_{\mathrm{\textit{T}}}(\mathcal{G},\mathrm{\textit{T}})-\\&
[2RR_{\zeta\eta}-4R^{\mu}_{\zeta}R_{\mu\eta}-4R_{\zeta\mu\eta\nu}R^{\mu\nu}+2R^{\mu\nu\delta}_{\zeta}R_{\eta\mu\nu\delta}]
f_{\mathcal{G}}(\mathcal{G},\mathrm{T})+\kappa^{2}\mathrm{\textit{T}}_{\zeta\eta},
\end{split}
\end{align}
where the symbols involved have their usual meanings and $\Theta_{\zeta\eta}= g^{\mu\nu}\frac{\delta
\mathrm{\textit{T}}_{\mu\nu}}{\delta g_{\zeta\eta}}$ and the subscript $\mathcal{G}$ or $\textit{T}$ in the functions denote the partial derivatives. It would be interesting to notice that if we substitute $f(\mathcal{G},\mathrm{\textit{T}})=f(\mathcal{G})$ in Eq.($\ref{4_eqn}$), then the field equations of $f(\mathcal{G})$ gravity are recovered. Moreover, the case $f(\mathcal{G}, \mathrm{\textit{T}})=0$ reduce the modified field equations to the usual GR equations.
For the sake of simplicity, from now onwards we consider $f(\mathcal{G},\mathrm{\textit{T}})\equiv f$,
$f_{\mathcal{G}}(\mathcal{G},\mathrm{\textit{T}})\equiv f_{\mathcal{G}}$ etc. The trace of Eq.($\ref{4_eqn}$) gives
\begin{equation}\label{5_eqn}
R+\kappa^{2}\mathrm{\textit{T}}-(\mathrm{\textit{T}}+\Theta)f_{\mathrm{\textit{T}}}+2f+2\mathcal{G}f_{\mathcal{G}}-2R\nabla^{2}f_{\mathcal{G}}
+4R^{\zeta\eta}\nabla_{\zeta}\nabla_{\eta}f_{\mathcal{G}}=0.
\end{equation}
It may be noticed that this relates $R$, $\mathcal{G}$ and $\mathrm{\textit{T}}$ differentially and not algebraically as in GR, where
$R=-\kappa\mathrm{\textit{T}}$. This indicates that the modified field equations may admit many solutions than other modified theories and GR.
The covariant divergence of Eq.(\ref{4_eqn}) is given by
\begin{equation}\label{div}
\nabla^{\zeta}T_{\zeta\eta}=\frac{f_{\mathrm{\textit{T}}}}
{\kappa^{2}-f_{\mathrm{\textit{T}}}}\bigg[(\mathrm{\textit{T}}_{\zeta\eta}+\Theta_{\zeta\eta})
\nabla^{\zeta}(\text{ln}f_{\mathrm{\textit{T}}})+
\nabla^{\zeta}\Theta_{\zeta\eta}-
\frac{g_{\zeta\eta}}{2}\nabla^{\zeta}\textit{T}\bigg],
\end{equation}
which is not zero. It is due to the presence of higher order derivatives of the energy momentum tensor that are naturally present in the field equations. Thus the theory might be plagued by divergences at astrophysical scales. This seems to be an issue with some other higher order derivatives theories as well that includes higher order terms of energy momentum tensor. However, to deal with the issue, one can put some constraints to Eq.(\ref{div}) to obtain standard conservation equation \cite{sharif.ayesha}. Here we take the spatially homogeneous, anisotropic, LRS Bianchi type $I$ spacetime
\begin{equation}\label{4}
ds^{2}=dt^2-X^2(t)dx^2-Y^2(t)[dy^2+dz^2],
\end{equation}
where $X$ and $Y$ are cosmic scale factors. Moreover, we consider that the universe is composed of perfect fluid
\begin{equation}\label{7}
\textit{T}_{\mu\nu}=(\rho + p)u_\mu u_\nu-pg_{\mu\nu},
\end{equation}
where $\rho$ and $p$ denote the energy density and pressure of the fluid respectively.
The Ricci scalar and GB invariant for Eq.(\ref{4}) is given as
\begin{equation}\label{5}
R=-2\bigg[\frac{\ddot{X}}{X}+2\frac{\ddot{Y}}{Y}+
\frac{2\dot{X}\dot{Y}}{XY}+\frac{\dot{Y}^2}{Y^2}\bigg],~~~
\mathcal{G}=8\bigg[\frac{\ddot{X}\dot{Y}^2}{XY^2}+2\frac{\dot{X}\dot{Y}\ddot{Y}}{XY^2}\bigg],
\end{equation}
where the overdot denotes the derivative with respect to the time coordinate.
Now we define some textbook physical quantities. The average scale factor $a$ and
average Hubble parameter $H$ for the model under consideration take the form
\begin{equation}\label{10}
a=\sqrt[3]{XY^2},\quad H=\frac{1}{3}(\frac{\dot{X}}{X}+\frac{2\dot{Y}}{Y}).
\end{equation}
The expansion scalar $\theta$ and shear scalar $\sigma$ are given as follows
\begin{eqnarray}\label{12}
\theta&=&u^\mu_{;\mu}=\frac{\dot{X}}{X}+2\frac{\dot{Y}}{Y},\\
\label{13} \sigma^2&=&\frac{1}{2}\sigma_{\mu\nu}\sigma^{\mu\nu}
=\frac{1}{3}[\frac{\dot{X}}{X}-\frac{\dot{Y}}{Y}]^2,
\end{eqnarray}
where
\begin{equation}\label{14}
\sigma_{\mu\nu}=\frac{1}{2}(u_{\mu;\alpha}h^\alpha_\nu+u_{\nu;\alpha}h^\alpha_\mu)
-\frac{1}{3}\theta h_{\mu\nu},
\end{equation}
$h_{\mu\nu}=g_{\mu\nu}-u_{\mu}u_{\nu}$ is the projection tensor.
Now for LRS Bianchi type $I$ spacetime (\ref{4}), the field equations (\ref{4_eqn}) take the form
\begin{eqnarray} \label{15}
2\bigg(2\frac{\dot{X}\dot{Y}}{XY}+\frac{\dot{Y}^2}{Y^2}\bigg)-24\frac{\dot{X}\dot{Y}^2}{XY^2}\dot{{f_\mathcal{G}}}+\mathcal{G}f_\mathcal{G}-f-2(\rho+p)f_\textit{T}=2\kappa^2\rho,\\\label{16}
-2\bigg(2\frac{\ddot{Y}}{Y}+\frac{\dot{Y}^2}{Y^2}\bigg)+16\frac{\dot{Y}\ddot{Y}}{Y^2}\dot{{f_\mathcal{G}}}+8\frac{\dot{Y}^2}{Y^2}\ddot{{f_\mathcal{G}}}-\mathcal{G}f_\mathcal{G}+f=2\kappa^2 p,\\\label{17}
-2\bigg(\frac{\ddot{X}}{X}+\frac{\ddot{Y}}{Y}+\frac{\dot{X}\dot{Y}}{XY}\bigg)+8(\frac{\dot{X}\ddot{Y}}{XY}+\frac{\dot{Y}\ddot{X}}{YX})\dot{{f_\mathcal{G}}}+
8\frac{\dot{X}\dot{Y}}{XY}\ddot{{f_\mathcal{G}}}-\mathcal{G}f_\mathcal{G}+f=2\kappa^2 p.
\end{eqnarray}
These are three highly non-linear and difficult differential equations with five unknowns. Thus we need an additional constraint to investigate any exact solution. Here we may consider a physical condition that shear scalar $\sigma$ is
proportional to expansion scalar $\theta$ which provides
\begin{equation}\label{18}
X=Y^n,
\end{equation}
where $n$ is an arbitrary real number.
In literature \cite{21}-\cite{Phyreason7}, many authors explored the exact solutions of field equations using this condition.
Thus using Eq.(\ref{18}), field equations (\ref{15})-(\ref{17}) take the form
\begin{eqnarray} \label{19}
&&2(2n+1)\frac{\dot{Y}^2}{Y^2}-24n\frac{\dot{Y}^3}{Y^3}\dot{{f_\mathcal{G}}}+\mathcal{G}f_\mathcal{G}-f-2(\rho+p)f_\textit{T}=2\kappa^2\rho,\\\label{20}
&&-2\bigg(2\frac{\ddot{Y}}{Y}+\frac{\dot{Y}^2}{Y^2}\bigg)+16\frac{\dot{Y}\ddot{Y}}{Y^2}\dot{{f_\mathcal{G}}}+8\frac{\dot{Y}^2}{Y^2}\ddot{{f_\mathcal{G}}}-\mathcal{G}f_\mathcal{G}+f=2\kappa^2 p,\\\nonumber
&&-2\bigg((n+1)\frac{\ddot{Y}}{Y}+n^2\frac{\dot{Y}^2}{Y^2}\bigg)+8\bigg(2n\frac{\dot{Y}\ddot{Y}}{Y^2}+n(n-1)\frac{\dot{Y}^3}{Y^3}\bigg)\dot{{f_\mathcal{G}}}+\\\label{21}
&&~~~~~~~~~~~~~~~~~~~~~~~~~~~~~~~~~~~~~~8n\frac{\dot{Y}^2}{Y^2}\ddot{{f_\mathcal{G}}}-\mathcal{G}f_\mathcal{G}+f=2\kappa^2 p.
\end{eqnarray}
Now we investigate the exact solutions of these field equations. %We consider two classes of

\section{Exact Solutions of Modified Field Equations}

We consider the $f(\mathcal{G},\textit{T})$ model as
\begin{eqnarray}\label{f(GT)}
f(\mathcal{G},\textit{T})=\alpha f_1(\mathcal{G})+ \beta f_2(\textit{T}),
\end{eqnarray}
where $\alpha$ and $\beta$ are arbitrary constants.
Further, we choose $f_1(\mathcal{G})$ in power law form, i.e.,
\begin{eqnarray}\label{f(G)}
f_1(\mathcal{G})=\mathcal{G}^{m+1}.
\end{eqnarray}
This model has already been proposed by Cognola et al. \cite{Cognola} and it is interesting
because the chances of appearing Big-Rip singularity vanish using this model.
Subtraction of Eqs. (\ref{20}) and (\ref{21}) yields
\begin{eqnarray} \label{22}
\frac{\ddot{Y}}{Y}+(n+1)\frac{\dot{Y}^2}{Y^2}-4\bigg(2\frac{\dot{Y}\ddot{Y}}{Y^2}+n\frac{\dot{Y}^3}{Y^3}\bigg)\dot{{f_\mathcal{G}}}-
4\frac{\dot{Y}^2}{Y^2}\ddot{{f_\mathcal{G}}}=0.
\end{eqnarray}
Using Eq.(\ref{f(G)}) in Eq.(\ref{f(GT)}), it follows that,
\begin{eqnarray}\label{f(GG)}
f_\mathcal{G}=\alpha(m+1)\mathcal{G}^{m},
\end{eqnarray}
For simplicity and without loss of any generality, we choose $\alpha=\frac{1}{m+1}$ so that Eq.(\ref{22}) takes the form
\begin{eqnarray} \label{23}
\frac{\ddot{Y}}{Y}+(n+1)\frac{\dot{Y}^2}{Y^2}-4m\mathcal{G}^m\bigg[\bigg(2\frac{\dot{Y}\ddot{Y}}{Y^2}+n\frac{\dot{Y}^3}{Y^3}\bigg)\frac{\dot{\mathcal{G}}}{\mathcal{G}}+
\frac{\dot{Y}^2}{Y^2}\bigg((m-1){\frac{\dot{\mathcal{G}}^2}{\mathcal{G}^2}}+\frac{\ddot{\mathcal{G}}}{\mathcal{G}}\bigg)\bigg]=0.
\end{eqnarray}
After inserting the value of GB invariant
%(\ref{6n}),
Eq.(\ref{23}) reduces to a differential equations with three unknowns $Y,~m$ and $n$.
It would be worthwhile to mention here that many solutions can be found using Eq.(\ref{23}). Here we consider the power law form, i. e.
\begin{equation}\label{24}
Y(t)=\gamma t^{k}
\end{equation}
where $\gamma$ and $k$ are arbitrary constants. Using this in Eq.(\ref{23}), we obtain a constraint equation
\begin{equation}\label{024}
t^{2+4m}(kn+2k-1)+16mk[8nk^3(kn+2k-3)]^m(kn+2k-4m-3)=0.
\end{equation}
This equation is satisfied for $k=\frac{1}{n+2}$ such that
\begin{equation}
m(2m+1)=0.
\end{equation}
Thus, corresponding to two roots of this equation, we obtain two choices of $f_1(\mathcal{G})$ models
\begin{equation}\label{25}
f_1(\mathcal{G})=\mathcal{G}+c_1,~~~~~~~~~~~~f_1(\mathcal{G})=2\sqrt{\mathcal{G}}+c_2,
\end{equation}
where $c_1$ and $c_2$ are integration constants. The first model recovers the usual GB gravity for $c_1=0$ and $f_2(\textit{T})=0$.
The second model with square root term is important as it leads to a viable inflation in the presence of massive
scalar field \cite{Myrzakul}. It is mentioned that different forms of $f_2(\textit{T})$ can be assumed to reconstruct the solutions. However,
we propose only two models for the present analysis.\\\\\\\\
\textbf{Case I: \textbf{Linear $f_2(\textit{T})$ Model}}\\\\
We consider linear form here for the sake of simplicity. Thus considering $f_2(\textit{T})=\textit{T}$ and using Eqs.(\ref{19})-(\ref{21}), the expression for energy density and pressure of universe turn out to be
\begin{eqnarray}\nonumber
\rho&=&\frac{1}{t^4(m+1)(n+2)^3\kappa^2(\kappa^2+1)}\bigg[t^2(m+1)(\kappa^2-1)(2n^2+5n+2)-\\\nonumber
&&8\bigg(\frac{-16n}{t^4(n+2)^3}\bigg)^m\big(4m^3(n+2)+m^2(n^2+10n-6\kappa^2n+12)+\\\label{rhon}
&&m(n^2+7n-5\kappa^2n+4)\big)\bigg]-\frac{\beta f_2(\textit{T})}{2\kappa^2},
\end{eqnarray}
\begin{eqnarray}\nonumber
p&=&\frac{1}{t^4(m+1)(n+2)^3\kappa^2}\bigg[t^2(m+1)(2n^2+5n+2)+8\bigg(\frac{-16n}{t^4(n+2)^3}\bigg)^m\\\nonumber
&&\big(4m^3(n+2)+m^2(n^2+10n+12)+m(n^2+7n+4)\big)\bigg]+\frac{\beta f_2(\textit{T})}{2\kappa^2}.\\\label{pn}
\end{eqnarray}
It is evident from Eqs.(\ref{rhon}) and (\ref{pn}) energy density and pressure of the
universe is defined for $-\infty<n<-2$ and $-2<n<0$ for $m=-1/2$.
Adding Eqs.(\ref{rhon}) and (\ref{pn}), we obtain
\begin{eqnarray}\label{rho+p}
\rho + p=\frac{2t^2(2n^2+5n+2)+
8\big(\frac{-16n}{t^4(n+2)^3}\big)^m\big(4m^2(n+2)+m(n^2+12n+4)\big)}{t^4(n+2)^3(\kappa^2+1)},
\end{eqnarray}
The behaviour of energy density plus pressure of universe can be seen from Fig.($1a$) for the model $f(\mathcal{G}, \mathrm{\textit{T}})=\sqrt{\mathcal{G}}+ \mathrm{\textit{T}}$ with $\kappa=1$. It is clear that $\rho+p\rightarrow 0$ as time grows which further suggests EoS parameter $\omega\rightarrow -1$. This is interesting as the phantom like dark energy is found to be in the region where $\omega<-1$. The universe with phantom dark energy ends up with a finite time future singularity known as cosmic doomsday or big rip \cite{Caldwell1,Caldwell2}. Moreover, accelerated expansion of universe is described when $\omega\approx-1$ \cite{nature1}-\cite{nature3}. Fig.($1b$) depicts the behavior of $\rho$
for radiation universe.\\\\\\\\\\
\textbf{Case II: \textbf{Starobinsky Like $f_2(\textit{T})$ Model}}\\\\
Here we propose $f_2(\textit{T})=\textit{T}+\epsilon \textit{T}^2$, where $\epsilon$ is an arbitrary constant.
For this model, manipulation of Eqs.(\ref{19})-(\ref{21}) yields,
\begin{eqnarray}\nonumber
&&(\rho+p)(1+\kappa^2+2\epsilon\rho-6\epsilon p)=\frac{2}{t^4(n+2)^3}\bigg[t^2(2n^2+5n+2)+\\
&&8\bigg(\frac{-16n}{t^4(n+2)^3}\bigg)^m\bigg(2m^2(n^2+3n+2)+m(n^2+7n+2)\bigg)\bigg].
\end{eqnarray}
Using EoS parameter $p=\omega\rho$, it follows that
\begin{eqnarray}\nonumber
&&2\epsilon(1-2\omega-3\omega^2)\rho^2+(1+\kappa^2)(1+\omega)\rho-\frac{2}{t^4(n+2)^3}\bigg[t^2(2n^2+5n+2)+\\\label{aanu}
&&8\bigg(\frac{-16n}{t^4(n+2)^3}\bigg)^m\bigg(2m^2(n^2+3n+2)+m(n^2+7n+2)\bigg)\bigg]=0.
\end{eqnarray}
Using Eq.(\ref{aanu}), we may analyze the behavior of universe by choosing different values
of EoS parameter.\\\\
\begin{itemize}
\item $\omega=0$ (Dust fluid)\\\\
In this case, the energy density of the universe turn out to be
\begin{eqnarray}\nonumber
\rho=\frac{-(1+\kappa^2)\pm\sqrt{(1+\kappa^2)^2+4l}}{2\epsilon},
\end{eqnarray}
where
\begin{eqnarray}\nonumber
l=\frac{2}{t^4(n+2)^3}\bigg[t^2(2n^2+5n+2)+
8\big(\frac{-16n}{t^4(n+2)^3}\big)^m\big(2m^2(n^2+3n+2)+m(n^2+7n+2)\big)\bigg].
\end{eqnarray}
The graphical behavior of energy density for both the roots is shown in Fig.($2a$) and ($2b$). The first root gives positive energy density but it approaches to zero as $t$ grows. However, the second root provides unrealistic behavior as energy density approaches to $-2$ at later times.
\item $\omega=1/4$ (Sub-relativistic fluid)\\\\
Here the energy density of the universe turn out to be
\begin{eqnarray}\nonumber
\rho=\frac{-10(1+\kappa^2)\pm2\sqrt{25(1+\kappa^2)^2+20l}}{5\epsilon}.
\end{eqnarray}
In this case, the first root Fig.($3a$) gives positive energy density having similar limiting value as in the dust case. However, the second root Fig.($3b$) gives negative energy density which corresponds to negative pressure universe. It is interesting because it is believed that dark energy has negative pressure causing accelerated expansion of universe.
\item $\omega=1/3$ (Radiation fluid)\\\\
In this case, Eq.(\ref{aanu}) becomes linear and the energy density of the universe turns out to be
\begin{eqnarray}\nonumber
\rho=\frac{3l}{4(1+\kappa^2)}.
\end{eqnarray}
The graphical behavior of energy density in this case is shown in ($1b$).
\item $\omega=1/2$ (Ultra-relativistic fluid)\\\\
Here the energy density of the universe turn out to be
\begin{eqnarray}\nonumber
\rho=\frac{3(1+\kappa^2)\mp\sqrt{(1+\kappa^2)^2-12l}}{3\epsilon}.
\end{eqnarray}
Here both the roots give positive energy density. However, the first root shows a decreasing behavior Fig.($4a$) while the send root gives increasing trend as the time passes Fig.($4b$).
\item $\omega=1$ (Stiff fluid)\\\\
Here the energy density of the universe turn out to be
\begin{eqnarray}\nonumber
\rho=\frac{(1+\kappa^2)\mp\sqrt{(1+\kappa^2)^2-4l}}{4\epsilon}.
\end{eqnarray}
The graphical behavior in this case is similar to that of ultra-relativistic fluid universe Fig.($5a$) and ($5b$).
\end{itemize}
\begin{figure}\center
\begin{tabular}{cccc}
\epsfig{file=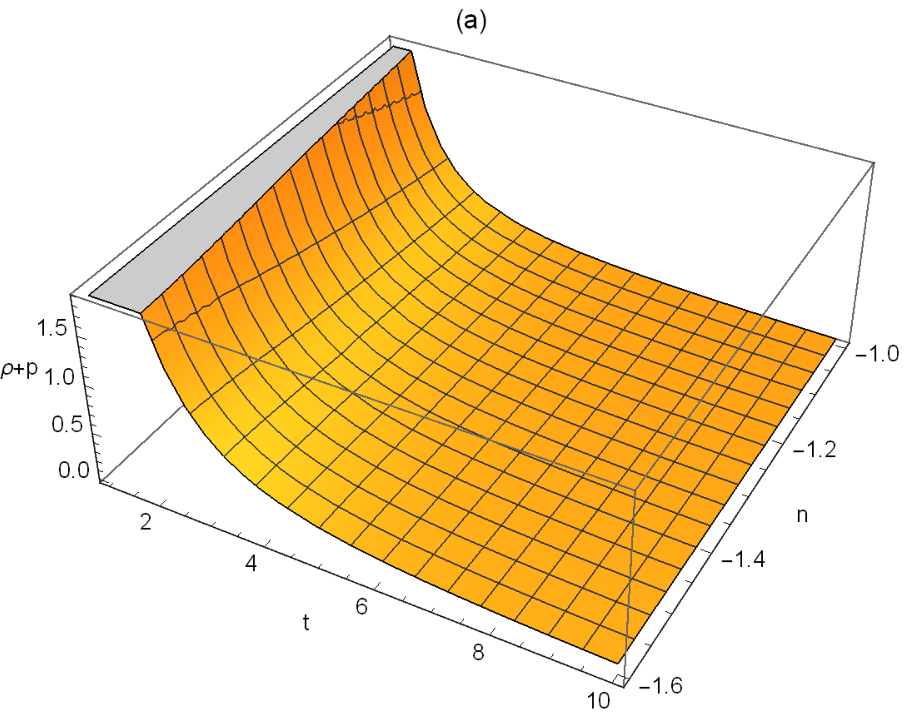,width=0.4\linewidth}&
\epsfig{file=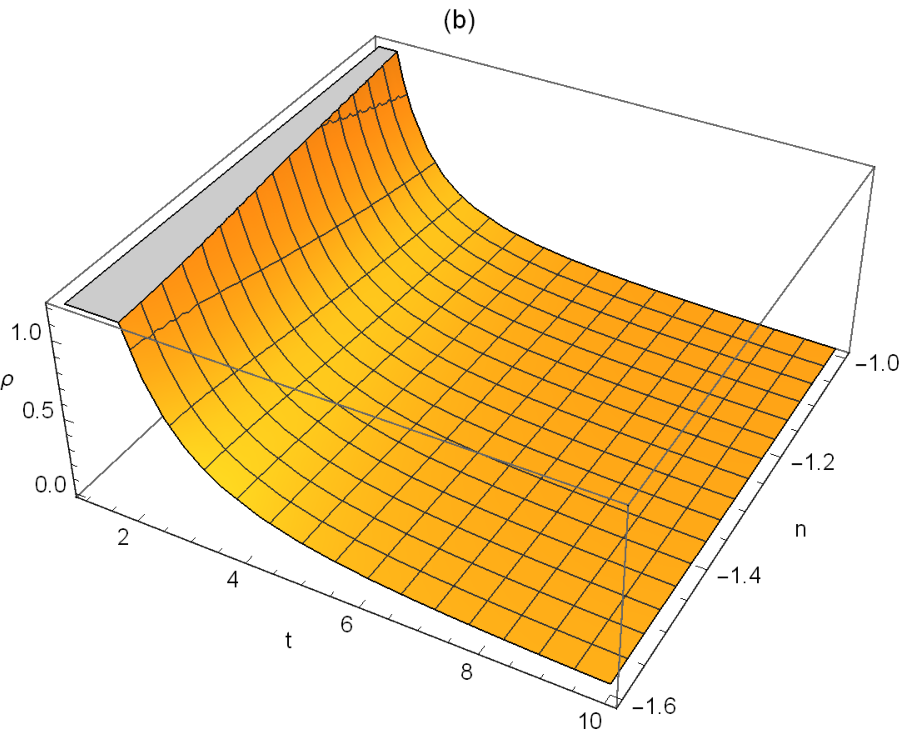,width=0.4\linewidth} \\
\end{tabular}
\caption{(a) Behavior of $\rho+p$ for $m=-1/2$ with $\kappa=1$, (b) Behavior of $\rho$ for Radiation universe.}\center
\end{figure}
\begin{figure}\center
\begin{tabular}{cccc}
\epsfig{file=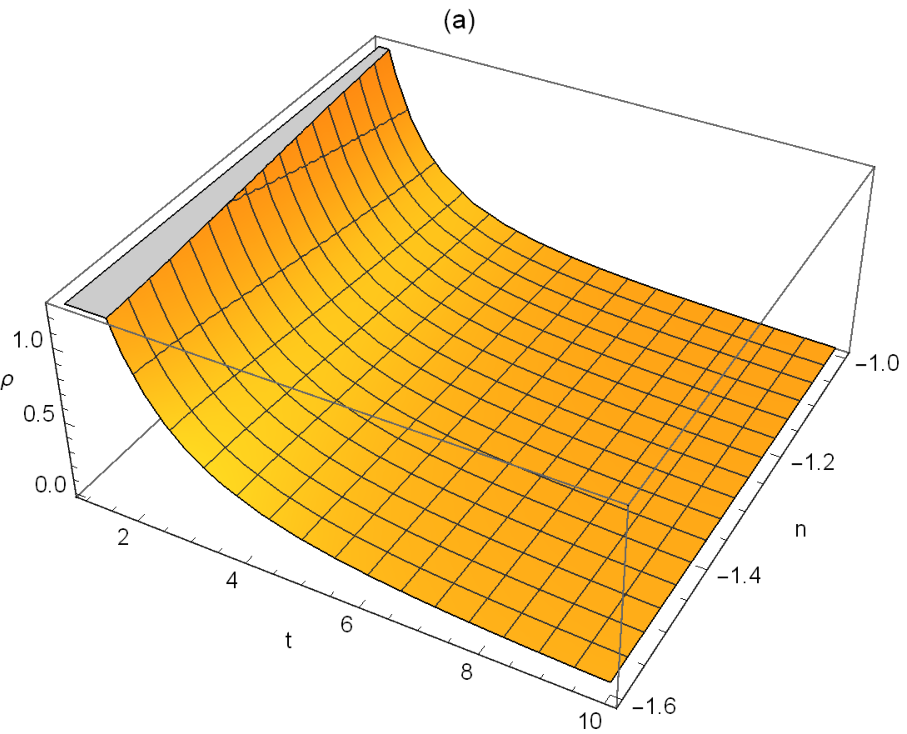,width=0.4\linewidth} &
\epsfig{file=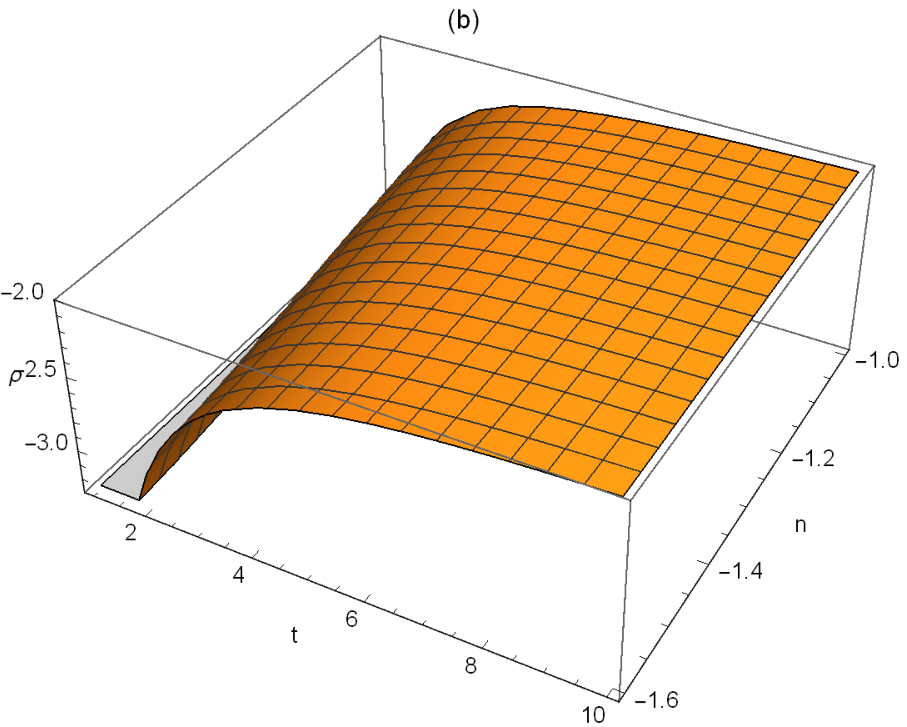,width=0.4\linewidth} \\
\end{tabular}
\caption{Behaviour of $\rho$ for Dust universe}\center
\end{figure}
\begin{figure}\center
\begin{tabular}{cccc}
\epsfig{file=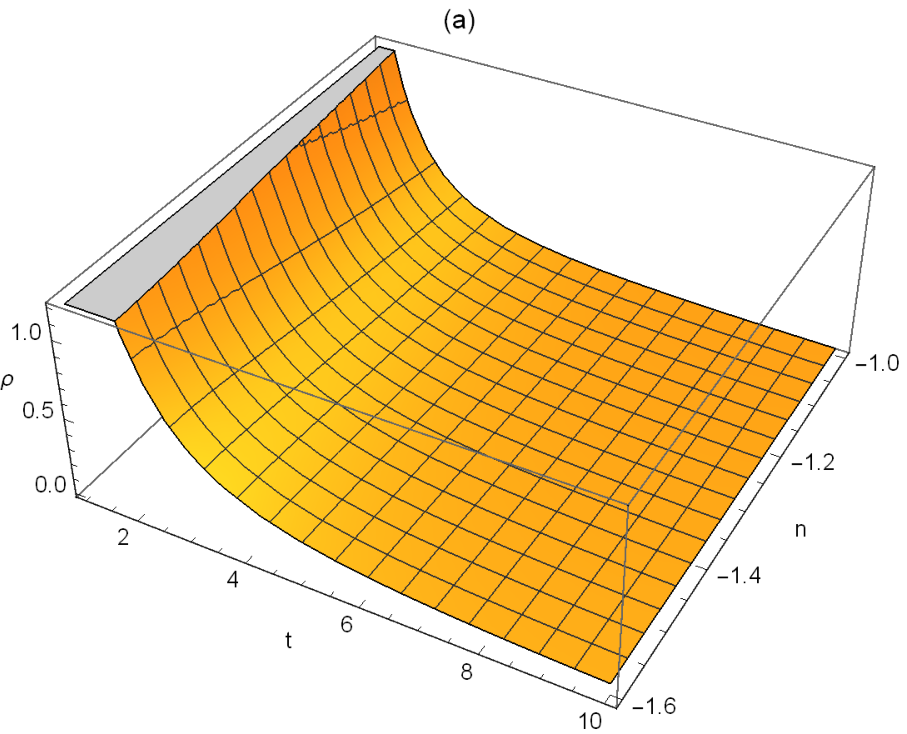,width=0.4\linewidth} &
\epsfig{file=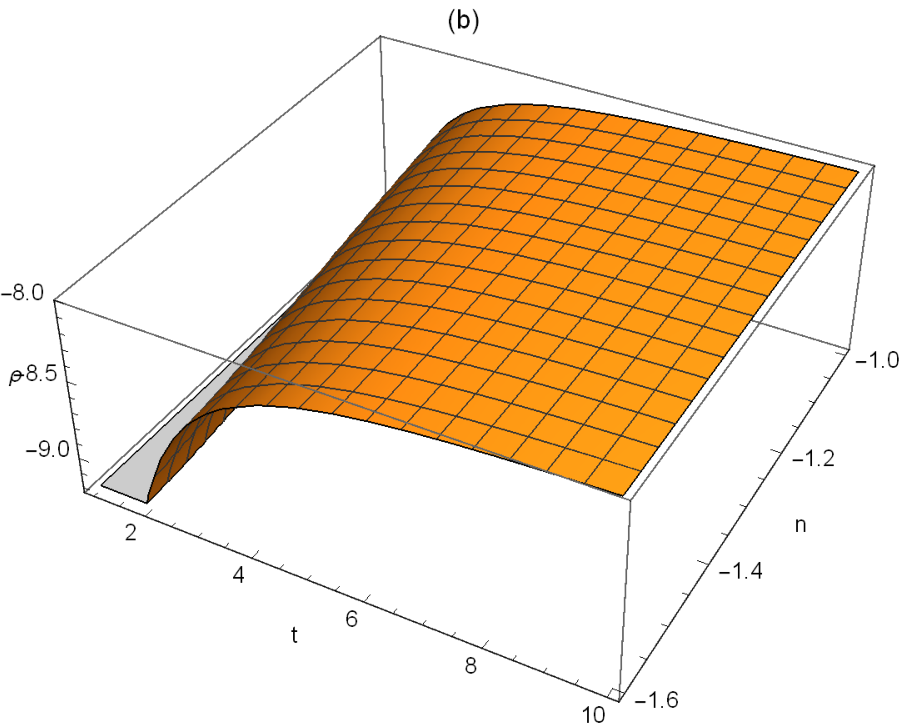,width=0.4\linewidth} \\
\end{tabular}
\caption{Behaviour of $\rho$ for Sub-relativistic universe}\center
\end{figure}
\begin{figure}\center
\begin{tabular}{cccc}
\epsfig{file=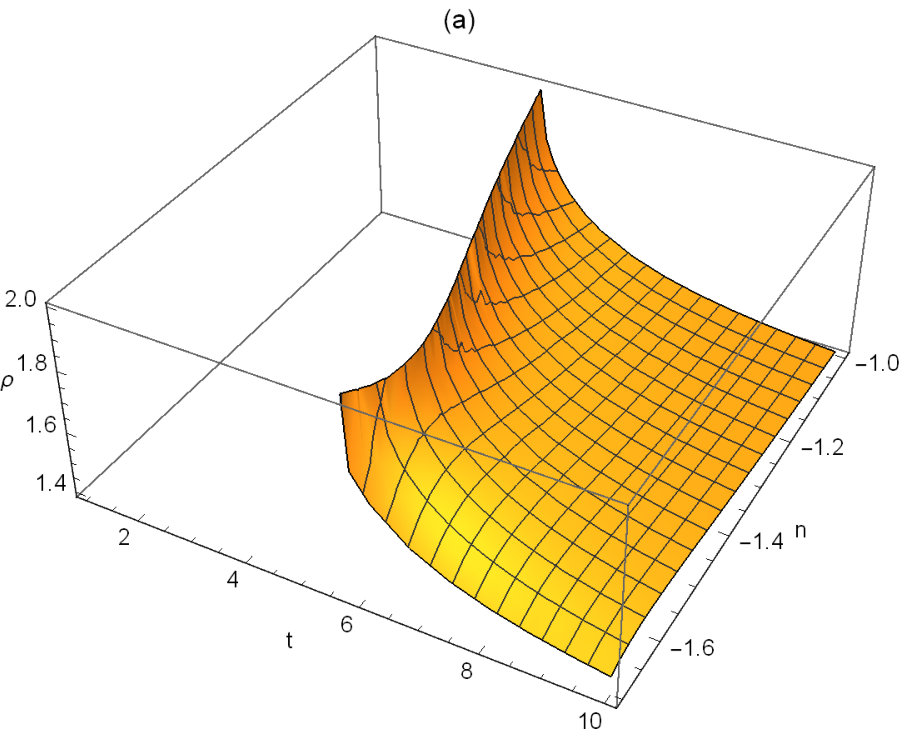,width=0.4\linewidth} &
\epsfig{file=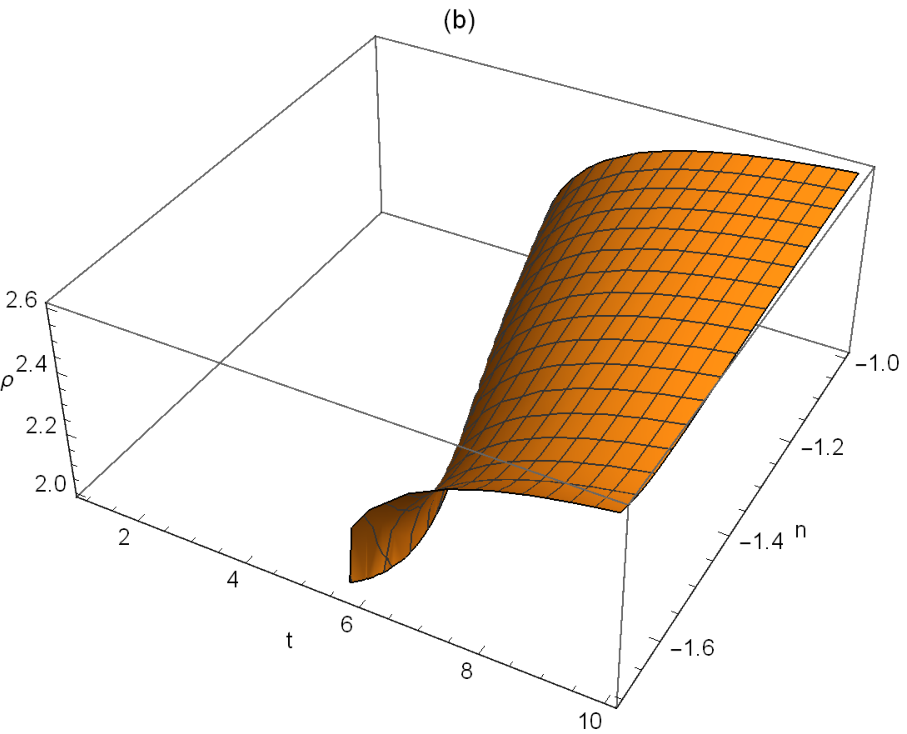,width=0.4\linewidth} \\
\end{tabular}
\caption{Behaviour of $\rho$ for Ultra-relativistic universe}\center
\end{figure}
\begin{figure}\center
\begin{tabular}{cccc}
\epsfig{file=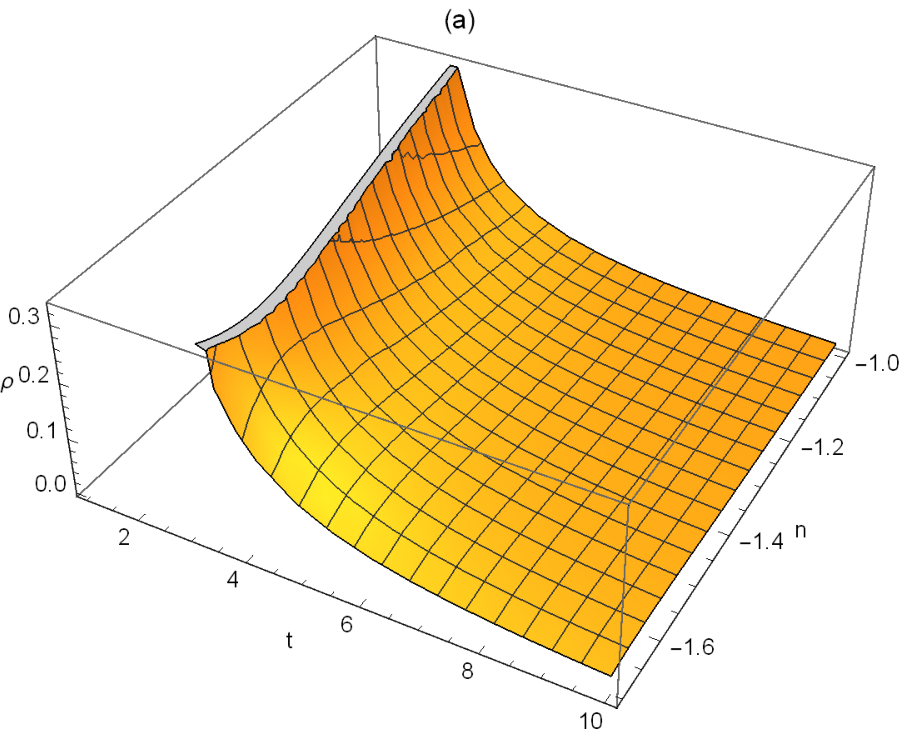,width=0.4\linewidth} &
\epsfig{file=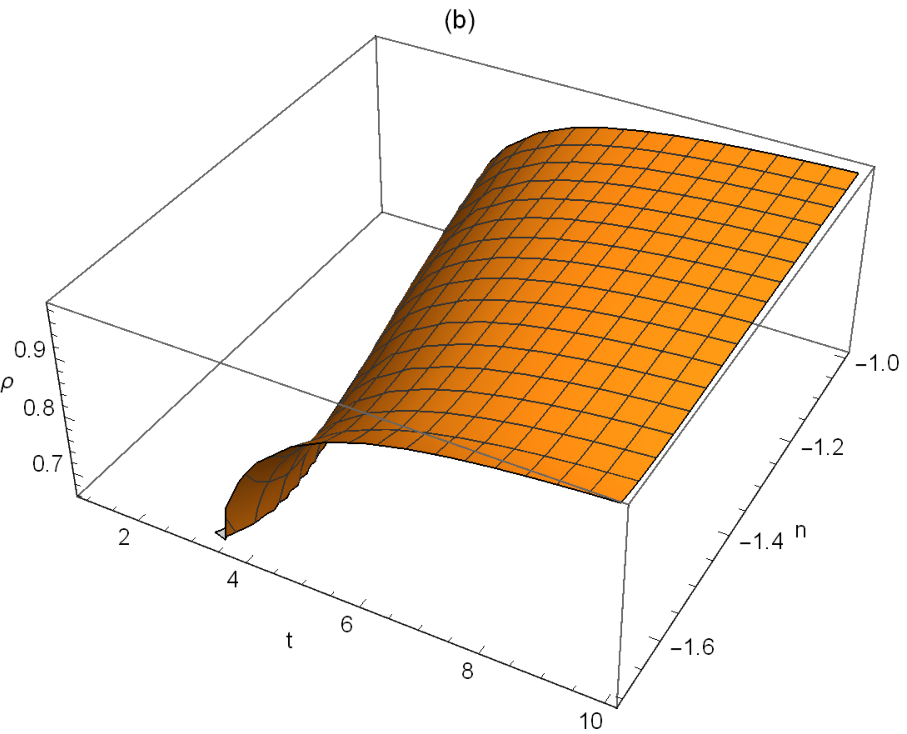,width=0.4\linewidth} \\
\end{tabular}
\caption{Behaviour of $\rho$ for Stiff universe}\center
\end{figure}
Thus, the solution metric takes the form
\begin{equation}\label{spacetime}
ds^{2}=dt^{2}-\gamma^{2n}t^{\frac{2n}{n+2}}d{x}^{2}-\gamma^2t^{\frac{2}{n+2}}(d{y}^{2}+d{z}^{2}).
\end{equation}
Thus many physical solutions with different values of EoS parameter are possible. It is due to the fact that the presence of term $\textit{T}$ in the bivariate function $f(\mathcal{G},\textit{T})$ may give many constraints which ultimately yield different solutions of the field equations and the theory may support the accelerated expansion of universe under certain conditions for the model under consideration.

\subsection{Discussion of Some Important Cosmological Parameters}

The Ricci scalar and GB invariant for the solution metric (\ref{spacetime}) takes the form
\begin{equation}
R=\frac{2(2n+1)}{(n+2)^2t^2},~~~~~~~~~~~~~\mathcal{G}=-\frac{16n}{(n+2)^3t^4}.
\end{equation}
It is obvious that the singularities exist at $t=0$ and $n=-2$.
The average Hubble parameter, average scale factor and volume scale factor of universe take the form
\begin{equation}\label{652910}
H=\frac{1}{3t},~~~~~~~~~~~~~a=\gamma^{\frac{n+2}{3}}t^{\frac{1}{3}},~~~~~~~~~~V=\gamma^{n+2}t.
\end{equation}
These parameters attain either zero value or tends to infinity at the points of singularity. In particular, the average scale factor is zero at the initial epoch $t=0$ and hence the model has a point type singularity \cite{MacCallum}.
The redshift for a distant source is directly related to the scale factor of the universe at the time when the photons were emitted from the source. The scale factor $a$ and redshift $z$ are related through the equation
\begin{equation}
a=\frac{a_0}{1+z},
\end{equation}
where $a_0$ is the present values of the scale factor.
Using Eq.(\ref{652910}), we get
\begin{equation}\label{22c290}
\frac{H}{H_0}=\frac{t_0}{t},~~~~\frac{a_0}{a}=1+z=\bigg(\frac{t_0}{t}\bigg)^{\frac{1}{3}},
\end{equation}
where $H_0$ represents the present values of Hubble's parameter.
We can write the value of Hubble's parameter in terms of redshift parameter as
\begin{equation}\label{refree1}
H=H_0(1+z)^3.
\end{equation}
The deceleration, jerk and snap parameters for the solution are given by
\begin{equation}
q=2,~~~~~~~~~j=10,~~~~~~~~~~s=-80.
\end{equation}
The expansion scalar and shear scalar become
\begin{equation}\label{22b}
\theta=\frac{1}{t},\quad \sigma^2=\frac{1}{3}\bigg[\frac{(n-1)}{(n+2)t}\bigg]^2.
\end{equation}
It is mentioned here that the isotropy condition $\frac{\sigma^2}{\theta}\rightarrow 0$ as $t\rightarrow \infty$,
is also satisfied in this case. It is also evident that $\frac{\sigma^2}{\theta}\rightarrow 0$ for small values of $n$ even when $t$ is not very large. This indicates that transition to isotropy is also possible for some suitable values of $n$ and $t$ (other than $\infty$). Thus, anisotropic universe isotropizes in general and FRW limit does exist with power law $f(\mathcal{G},\textit{T})$ model under consideration.
It can be seen from Eqs. (\ref{652910}) and (\ref{22b}) that the volume of universe is zero at $t=0$ while the expansion scalar is infinite, which suggests that the universe started it evolution with zero volume at $t=0$, i.e. big bang scenario.

\section{Reconstruction of Some Important Cosmological Solutions}

Here we discuss some special cases to reconstruct some important cosmological solutions.

\subsection{Flat FRW Solution}

For a special case when $n=1$, and $m=-\frac{1}{2}$, space-time (\ref{spacetime}) takes the form
\begin{equation}\label{spacetimefrw}
ds^{2}=dt^{2}-\gamma^{2}t^{\frac{2}{3}}(d{x}^{2}+d{y}^{2}+d{z}^{2}),
\end{equation}
which is the solution of well-known flat FRW metric. In this case $\omega$ attains an imaginary value which shows that
the model $f(\mathcal{G},\textit{T})=\sqrt{\mathcal{G}}+ \beta f_2(\textit{T})$ is not viable for flat FRW solution. However, when $m=0$,
using Eqs.(\ref{rhon}) and (\ref{pn}), EoS parameter turns out to be
\begin{equation}
\omega=\frac{(1+\kappa^2)\big(2+3\beta f_2(\textit{T})t^2\big)}{3\beta f_2(\textit{T})t^2(1+\kappa^2)+2(1-\kappa^2)}.
\end{equation}
One can further analyze using different possibilities $f_2(\textit{T})$ and hence in this case the reconstruction flat FRW solution is justified.

\subsection{ Kasner Type Solution}

We can recover interesting Kasner type solution by putting $n=-{1/2}$ in Eq.(\ref{spacetime})
\begin{equation}
ds^{2}=dt^{2}-\gamma^{-1}t^{-\frac{2}{3}}d{x}^{2}-\gamma^{2}t^{\frac{4}{3}}({y}^{2}+d{z}^{2}).
\end{equation}
It is similar to the well-known Kasner's metric \cite{30} and one can obtain the exact spacetime after redefining the parameters.
Here EoS parameter takes the form
\begin{equation}\label{29919}
\omega=-\frac{(1+\kappa^2)\big[64m^2(7+6m)\big(\frac{1728}{t^4}\big)^m+27\beta f_2(\textit{T})t^4(m+1)729^m\big]}{64m(6m\kappa^2+5\kappa^2+7m+6m^2)\big(\frac{1728}{t^4}\big)^m+27\beta f_2(\textit{T})t^4(m+1)(1+\kappa^2)729^m}.
\end{equation}
It is interesting to notice that for $m=0$, Eq.(\ref{29919}) gives $\omega=-1$ independent of the choice of $f_2(\textit{T})$ model and thus describing accelerated expansion of universe \cite{nature1,nature2,nature3}. Moreover, when $m=-1/2$, EoS parameter turns out to be
\begin{equation}\label{299190}
\omega=\frac{(1+\kappa^2)\big(16\sqrt{3}+9\beta f_2(\textit{T})t^4\big)}{16\sqrt{3}(\kappa^2-1)-9\beta f_2(\textit{T})t^4(1+\kappa^2)}.
\end{equation}
Hence our both values of $m$ incorporates Kasner type solution.

\subsection{Exponential Law Solutions}

It is to be noticed that the Eq.(\ref{23}) have an exponential solution of the form
\begin{equation}
Y(t)=e^{c_3t+c_4},
\end{equation}
with the constraint equation
\begin{equation}
n+2=0,
\end{equation}
where $c_3$ and $c_4$ are arbitrary constants. Here EoS parameter becomes
\begin{equation}
\omega=-\frac{(1+\kappa^2)(\beta f_2(\textit{T})-6)}{6(\kappa^2-1)+\beta f_2(\textit{T})(1+\kappa^2)},
\end{equation}
which is independent of parameter $m$ and hence any viable power law $f_1(\mathcal{G})$ gravity model can be used with an appropriate $f_2(\textit{T})$ model.
The solution metric in this case turn out to be
\begin{equation}
ds^{2}=dt^2-e^{-4(c_3t+c_4)}dx^2-e^{2(c_3t+c_4)}(dy^2+dz^2).
\end{equation}
The average Hubble parameter is be zero for this solution. All other dynamical
parameters expansion scalar $\theta$, shear scalar $\sigma$, volume scale factor of universe are constant here.

\section{Final Remarks}

The main purpose of this work is to investigate newly introduced modified GB theory namely $f(\mathcal{G},\textit{T})$ gravity. For this purpose, anisotropic background is chosen. LRS Bianchi type $I$ space-time is the most simplest model and has been used frequently in different contexts. To our knowledge, this is the first attempt to investigate the exact solutions of $f(\mathcal{G},\textit{T})$ gravity for LRS Bianchi type $I$ space-time. Moreover, we have not used any conventional assumption like constant deceleration parameter to investigate the exact solutions.
It is mentioned here that the field equations are highly nonlinear and it is due to the inclusion of bivariate function in the standard action.
So we assume that the shear scalar $\sigma$ proportional to the expansion scalar $\theta$ which gives $X=Y^n$, where
$X,~Y$ are the metric coefficients and $n$ is an arbitrary constant. We consider the model $f(\mathcal{G},\textit{T})=\alpha f_1(\mathcal{G})+ \beta f_2(\textit{T})$,
where $\alpha$ and $\beta$ are arbitrary constants.
Moreover, we have used a power law form $f_1(\mathcal{G})$ gravity model already available in literature \cite{Cognola}. The interesting feature of this model is that the chances of appearing Big-Rip singularity are minimized.
Further, the viability of this model has already been discussed in different cosmological contexts \cite{Model1,Model2,Model3}. We have also proposed a Starobinsky like $f_2(\textit{T})$ model, i. e. $f_2(\textit{T})=\textit{T}+\epsilon \textit{T}^2$, where $\epsilon$ is an arbitrary constant.
The interesting aspect of the Starobinsky like $f_2(\textit{T})$ model is that it provides many exact solutions and one can choose the best fit solutions according to the requirements.

Using the modified field equations, we have formulated a general differential equation (\ref{23}) which can further be used to investigate exact solutions.
Mainly we used power law and exponential forms of metric coefficients to explore the exact solutions of modified field equations.
It is shown that two $f_1(\mathcal{G})$ gravity models are associated with the power law solution. The first model recovers the usual GB gravity for $c_1=0$ and $f_2(\textit{T})=0$. However, the second model with square root term looks important as it leads to a viable inflation in the presence of massive
scalar field \cite{Myrzakul}. It is mentioned here that different forms of $f_2(\textit{T})$ can be assumed to reconstruct the solutions. However, we have discussed only two models for the present analysis. Firstly, we choose linear $f_2(\textit{T})$ gravity model. The expressions for energy density and pressure of universe are defined for for anisotropy parameter $-\infty<n<-2$ and $-2<n<0$ with $m=-1/2$. The graphical behavior (see fig.($1a$)) shows that $\rho+p\rightarrow 0$ as time grows which implies that EoS parameter $\omega\rightarrow -1$. This is interesting as the phantom like dark energy is found to be in the region where $\omega<-1$. The universe with phantom dark energy ends up with a finite time future singularity known as cosmic doomsday or big rip \cite{Caldwell1,Caldwell2}. Moreover, accelerated expansion of universe with de-Sitter type evolution is described when $\omega\approx-1$ \cite{nature1}-\cite{nature3}. However, the second model proposed as Starobinsky like model involves squared term of $\textit{T}$. This gives a quadratic equation in $\rho$ and we have analyzed the behavior of universe by choosing different values of EoS parameter. In particular, analysis is given for $\omega=0$ (Dust fluid), $\omega=1/4$ (Sub-relativistic fluid), $\omega=1/3$ (Radiation fluid), $\omega=1/2$ (Ultrarelativistic fluid) and $\omega=1$ (Stiff fluid). The graphical behavior of energy density for cases is discussed. In particular, in case of sub-relativistic fluid, one root corresponds to negative pressure universe. It is interesting because it is believed that dark energy has negative pressure causing accelerated expansion of universe. Thus, it is concluded that presence of term $\textit{T}$ in the bivariate function $f(\mathcal{G},\textit{T})$ may give many solutions of the field equations and the theory supports the accelerated expansion of universe under certain conditions for the model under consideration.

Lastly, we have reconstructed some important cosmological solutions. The first solution corresponds to flat FRW spacetime. This solution is valid for $\omega=1$ describing the stiff fluid universe for the first model. However, the solutions is not physical for the second model as $\omega$ is imaginary.
The second solution provides the well-known Kasner's universe and it gives the the value of anisotropy parameter $n=-1/2$. The third solution is obtained by exponential law assumption. It gives the average Hubble parameter zero and all other dynamical
parameters like expansion scalar $\theta$, shear scalar $\sigma$, volume scale factor of universe constant. EoS parameter is independent of the parameter $m$ for this solution and we get an explicit expression for $\omega$ involving $f_2(\textit{T})$. Thus any power law $f_1(\mathcal{G})$ gravity model may be used with an appropriate $f_2(\textit{T})$ model to reconstruct some important cosmological solutions.\\\\\\
\textbf{Conflict of Interest}\\\\The author declares that there is no conflict of interest regarding the publication of this paper.


\vspace{0.05cm}

\begin{thebibliography}{40}

\bibitem{5} Starobinsky, A.A.: Phys. Lett. \textbf{B91}(1980)99;
Starobinsky, A.A.: JETP Lett. \textbf{86}(2007)156;
Appleby, S.A. and Battye, R.A.: Phys. Lett. \textbf{B654}(2007)7;
Appleby, S.A., Battye, R.A. and Starobinsky, A.A.: JCAP \textbf{1006}(2010)005.

\bibitem{500} Nojiri, S. and Odintsov, S.D.: Phys. Rev. \textbf{D70}(2004)103522;
Harko, T., Lobo, F.S.N., Nojiri, S. and Odintsov, S.D.: Phys. Rev. \textbf{D84}(2011)024020;
Nojiri, S. and Odintsov, S.D.: Int. J. Geom. Meth. Mod. Phys. \textbf{115}(2007)4;
Bamba, K., Capozziello, S., Nojiri, S. and Odintsov, S.D.: Astrophys. Space Sci. \textbf{342}(2012)155;

\bibitem{27}Sharif, M. and Shamir, M.F.: Class. Quantum Grav.
\textbf{26}(2009)235020; Sharif, M. and Shamir, M.F.: Gen.
Relativ. Gravit. \textbf{42}(2010)2643.

\bibitem{501} Hu, W. and Sawicki, I.: Phys. Rev. \textbf{D76}(2007)064004;
Felice, A.D and Tsujikawa, S.:
Living Rev. Rel. \textbf{13}(2010)3; Sotiriou, T.P. and Faraoni,
V.: Rev. Mod. Phys. \textbf{82}(2010)451; Clifton, T., Ferreira,
P.G., Padilla, A. and Skordis, C.: Phys. Rept. \textbf{513}
(2012)1.

\bibitem{ft} Velten, H. and Carames, T.R.P.: Phys. Rev.
\textbf{D95}(2017)123536.

\bibitem{fG} Nojiri, S. and Odintsov, S.D.: Phys. Lett. \textbf{B631}(2005)1;
Nojiri, S., Odintsov, S.D. and Gorbunova, O.G.: J. Phys. A
\textbf{39}(2006)6627; Cognola, G., Elizalde, E., Nojiri, S., Odintsov, S.D.
and Zerbini, S.: Phys. Rev. D \textbf{73}(2006)084007.

\bibitem{Chiba} Chiba, T.: J. Cosmol. Astropart. Phys. \textbf{03}(2005)008.

\bibitem{sharif.ayesha} Sharif, M. and Ikram, A.: Eur. Phys. J. \textbf{C76}(2016)640.

\bibitem{abbas} Abbas, G., Momeni, D., Ali, M.A, Myrzakulov, R. and Qaisar, S.: Astrophys. Space Sci.
\textbf{357}(2015)158.

\bibitem{Houndjo1} Houndjo, M.J.S., Rodrigues, M.E., Momeni, D. and Myrzakulov, R.: Can. J. Phys. \textbf{92}(2014)1528.

\bibitem{Houndjo2} Rodrigues, M.E., Houndjo, M.J.S., Momeni, D. and Myrzakulov, R.: Can. J. Phys. \textbf{92}(2014)173.

\bibitem{fRG1} Wu, B. and Ma. B.: Phys. Rev. D \textbf{92}(2015)044012.

\bibitem{fRG2} Laurentis, M., Paolella, M. and Capozziello, S.: Phys. Rev. D \textbf{91}(2015)083531.

\bibitem{warm} Sharif, M. and Ikram, A.: JETP \textbf{123}(2016)40.	

\bibitem{noetherfG1} Sharif. M., Fatima. H.I.: JETP \textbf{122}(2016)104.

\bibitem{noetherfG2} Sharif. M., Fatima. H.I.: Int. J. Mod. Phys. D \textbf{25}(2016)1650011.

\bibitem{Garcia} Garcia, N.M., Harko, T., Lobo, F.S.N. and Mimoso, J.P.: J. Phys. Conf. Ser. \textbf{314}(2011)012060.

\bibitem{Atazadeh} Atazadeh, K. and Darabi, F.: Gen. Relativ. Gravit. \textbf{46}(2014)1664.

\bibitem{sharif.ayesha1} Sharif, M. and Ikram, A.: Physics of the Dark Universe \textbf{17}(2017)1.	

\bibitem{shamir.ahmad} Shamir, M. and Ahmad, M.: Eur. Phys. J. \textbf{C77}(2017)55.

\bibitem{sharif.ayesha2} Sharif, M. and Ikram, A.: arXiv:1707.05162.

\bibitem{shamir.ahmad1} Shamir, M. and Ahmad, M.: Mod. Phys. Lett. \textbf{A32}(2017)1750086.

\bibitem{21}Xing-Xiang, W.: Chin. Phys. Lett.
\textbf{22}(2005)29.

\bibitem{Phyreason3} Thorne, K.S.: Astrophys. J. \textbf{148}(1967)51.

\bibitem{Phyreason4} Collins, C.B. and Hawking, S.W.: Astrophys. J. \textbf{180}(1973)317.

\bibitem{Phyreason7} Sharif, M., Zubair, M.: Astrophys. Space Sci. \textbf{330}(2010)399.

\bibitem{Cognola} Cognola, G., Elizalde, E., Nojiri, S., Odintsov, S.D. and Zerbini, S.: Phys. Rev. D \textbf{73}(2006)084007.

\bibitem{Myrzakul} Myrzakul, S., Myrzakulov, R. and Sebastiani, L.: Eur. Phys. J. C \textbf{75}(2015)111.

\bibitem{Caldwell1} Starobinsky, A.A.: Grav. Cosmol. \textbf{6}(2000)157.

\bibitem{Caldwell2} Caldwell, R.R.: Phys. Lett. \textbf{B545}(2002)23.

\bibitem{nature1} Hogan, J.: Nature \textbf{448}(2007)240.

\bibitem{nature2} Corasaniti, P. S. et al: Phys. Rev. \textbf{D70}(2004)083006.

\bibitem{nature3} Weller, J., Lewis, A.M.: Mon. Not. Astron. Soc. \textbf{346}(2003)987.

\bibitem{MacCallum} MacCallum, M.A.H.: Commun. Math. Phys. \textbf{20}(1971)57.

\bibitem{stiff} Mathew, T.K., Aswathy, M.B. and Manoj, M.: Eur. Phys. J. C \textbf{74}(2014)3188.

\bibitem{30} Cataldo, M. and  Campo, S.D.: Phys. Rev.
\textbf{D61}(2000)128301.

\bibitem{Kim} Kim, W. and Yoon, M.S.: J. Korean Phys. Soc. \textbf{50}(2007)941.

\bibitem{Model1}  Cognola, G., Gastaldi, M. and Zerbini, S.: Int. J. Theor. Phys. \textbf{47}(2008)898.

\bibitem{Model2} Felice, A.D. and Tsujikawa, S.: Phys. Lett. B\textbf{675}(2009)1.

\bibitem{Model3} Bamba, K., Odintsov, S.D., Sebastiani, L. and Zerbini, S.: Eur. Phys. J. C\textbf{67}(2010)295.

\bibitem{Plank} Ade, P.A.R. et al.: Astron. Astrophys. \textbf{571}(2014)A16.
\bibitem{022}Wang, X.: Astrophys. Space
\bibitem{26}Paul, B.C., Debnath, P.S. and Ghose, S.: Phys. Rev.


\end{thebibliography}
\end{document}